\documentclass[10pt,journal,compsoc,review]{IEEEtran}

\usepackage{booktabs,multirow}  % For formal tables
\usepackage{amsfonts}           % math equation alignment
\usepackage{amsmath}            % blackboard math symbols
\usepackage{amsthm}
\usepackage{siunitx}
\usepackage{colortbl}       % colored tables
\usepackage{xcolor}   
\usepackage{graphicx}
\usepackage{float}
\usepackage[numbers,sort&compress]{natbib}
\usepackage{soul}
\usepackage{makecell}
\usepackage{booktabs}

\newcommand{\mc}[1]{\makecell{#1}}

\usepackage{hyperref}
\usepackage[capitalise]{cleveref}

\begin{document}

\title{Neural Cone Radiosity for Interactive Global Illumination with Glossy Materials}


% \renewcommand{\shortauthors}{Remi and Remies.}


\author{Jierui Ren, Haojie Jin, Bo Pang, Yisong Chen, Guoping Wang, Sheng Li*,~\IEEEmembership{Member,~IEEE} 
\IEEEcompsocitemizethanks{
\IEEEcompsocthanksitem Jierui Ren is with the College of Future Technology, Peking University, China. jerry@stu.pku.edu.cn
\IEEEcompsocthanksitem Haojie Jin, Bo Pang, Yisong Chen, Guoping Wang, and Sheng Li are with the School of Computer Science, Peking University, China.\\ E-mail: \ 
 \{jhj$|$chenyisong$|$wgp$|$lisheng\}@pku.edu.cn, bo98@stu.pku.edu.cn \\
Guoping Wang and Sheng Li are also with the National Key Laboratory of Intelligent Parallel Technology. \\
\IEEEcompsocthanksitem Sheng Li is the corresponding author.
}
}

\IEEEtitleabstractindextext{

%     \centering
%     \includegraphics[width=0.9\textwidth]{images/teaser.pdf}
%     \caption{\small Rendered results on two representative scenes. For each scene, we show the full-resolution image rendered with our method, along with close-up comparisons between the reference (path traced with 100{,}000 spp) and our result in selected regions. Our method achieves high visual fidelity while preserving fine details in both diffuse and glossy areas. Mean Absolute Percentage Error (MAPE) with respect to the reference is shown in the bottom-right corner of the whole image and each close-up comparison, providing a quantitative measure of reconstruction accuracy. }
%     \label{fig:teaser}

%    \centering
%    \includegraphics[width=0.9\textwidth]{images/teaser-split.pdf}
%    \caption{\small Side-by-side comparisons between our method and Neural Radiosity (NR) \cite{hadadan2021neural}. Our method achieves higher visual fidelity over the alternative approach in terms of Mean Absolute Percentage Error (MAPE). We also highlight the glossy material from highly glossy to various levels of glossiness.}
%    \label{fig:teaser}

\vspace{10pt}

\begin{abstract}
Modeling of high-frequency outgoing radiance distributions has long been a key challenge in rendering, particularly for glossy material. Such distributions concentrate radiative energy within a narrow lobe and are highly sensitive to changes in view direction. However, existing neural radiosity methods, which primarily rely on positional feature encoding, exhibit notable limitations in capturing these high-frequency, strongly view-dependent radiance distributions.
To address this, we propose a highly-efficient approach by reflectance-aware ray cone encoding based on the neural radiosity framework, named neural cone radiosity. The core idea is to employ a pre-filtered multi-resolution hash grid to accurately approximate the glossy BSDF lobe, embedding view-dependent reflectance characteristics directly into the encoding process through continuous spatial aggregation. Our design not only significantly improves the network’s ability to model high-frequency reflection distributions but also effectively handles surfaces with a wide range of glossiness levels, from highly glossy to low-gloss finishes. Meanwhile, our method reduces the network’s burden in fitting complex radiance distributions, allowing the overall architecture to remain compact and efficient.
Comprehensive experimental results demonstrate that our method consistently produces high-quality, noise-free renderings in real time under various glossiness conditions, and delivers superior fidelity and realism compared to baseline approaches.

\end{abstract}

%%
%% Copy pasete generated code by the tool at http://dl.acm.org/ccs.cfm.


}

\maketitle
\IEEEpeerreviewmaketitle

\begin{figure*}
    \centering
    \includegraphics[width=0.9\textwidth]{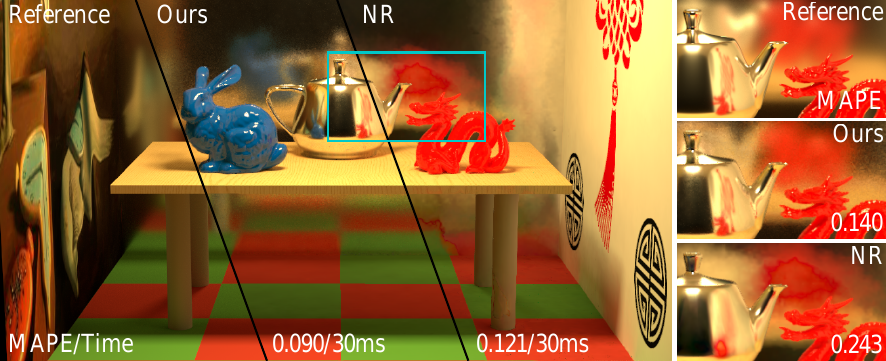}
    \caption{\small Side-by-side comparisons between our method and Neural Radiosity (NR) \cite{hadadan2021neural}. Our method achieves higher visual fidelity over the alternative approach in terms of Mean Absolute Percentage Error (MAPE). We also highlight the glossy material from highly glossy to various levels of glossiness.}
    \label{fig:teaser}
\end{figure*}

\section{Introduction}

Global illumination is a fundamental problem in computer graphics and plays a crucial role in physically-based rendering. Monte Carlo-based methods, such as path tracing, are standard solutions for producing high-quality global illumination. However, their high computational cost makes them impractical for real-time or interactive applications. 

%With the advent of deep learning, many recent approaches have leveraged neural networks to address this challenge. Some methods employ neural networks to denoise low-sample-per-pixel outputs from path tracers, while others directly predict indirect illumination effects using geometric buffers and cheap-to-compute direct lighting. Beyond these image-space techniques, neural networks have also been used to learn complex 3D spatial distributions that are difficult to model with traditional data structures.

Recent advances in deep learning have spurred the development of Neural Global Illumination (NGI) methods, which leverage the representational power of neural networks to approximate or accelerate GI computation. These methods can be broadly categorized into three classes: image-space denoising, which reconstructs high-quality images from sparse Monte Carlo samples using spatio-temporal filtering and learned priors \cite{Chaitanya17,Bako17,Bitterli18}; hybrid prediction methods, which combine geometric buffers with inexpensive direct lighting to predict indirect lighting \cite{Zhang20, Chaitanya17,nalbach2017deep,xin2021lightweight,isola2017image}; and 3D neural representations, which store and query radiance or irradiance directly in continuous 3D space \cite{muller2021real, hadadan2021neural}.

While NGI techniques have made remarkable progress, they still struggle to capture high-frequency outgoing radiance distributions. Such features, including glossy reflections, sharp highlights, and caustics, are notoriously difficult to reconstruct under low-sample-per-pixel budgets. Among these, glossy materials remain the most prevalent and challenging case. Glossy BRDF lobes exhibit narrow, view-dependent energy distributions that require high-quality sampling to avoid noise, but aggressive filtering often leads to over-blurring and bias \cite{Akerlund23, Heidrich09}. Probe- or cache-based GI systems handle such cases more stably but tend to lose sharpness in glossy regions \cite{Akerlund23}. Neural approaches that rely on secondary ray queries (e.g., Neural Radiance Caching \cite{muller2021real}) in glossy scenarios still face the noise–bias trade-off, while directly evaluating a neural radiance field at the primary hit point often results in overly smoothed reflections \cite{muller2021real, hadadan2021neural}. As noted, glossy surfaces remain a “hotspot” failure case for many real-time GI pipelines.

%NR follows a two-stage pipeline comprising training and rendering. During training, surface interactions are randomly sampled, and their outgoing radiance is supervised via Monte Carlo integration over incident directions. Since each incident direction leads to another surface interaction, the supervision is self-contained and does not require expensive pre-rendered ground truth.
As the representative, Neural Radiosity (NR) \cite{hadadan2021neural} can produce high-quality, noise-free renderings at interactive frame rates. 
It employs a neural representation over 3D space to model a scene’s outgoing radiance distribution and can capture certain high-frequency spatial details.
However, NR still struggles with complex directional distributions, particularly on glossy surfaces such as frosted mirrors or polished metals. Unlike perfect specular reflections, which can be resolved by tracing rays until they hit a non-specular surface, glossy reflections require integrating incident radiance modulated by a BSDF lobe. In RHS-style NR and Neural Radiance Caching \cite{muller2021real}, tracing a secondary ray and querying the network may introduce significant noise, whereas evaluating the network directly at the primary intersection often results in substantial bias and overly blurred reflections (see Figure \ref{fig:compare}).

To address this limitation in handling a wide range of glossy materials, we propose Neural Cone Radiosity (NCR). The core idea is to use ray cone encoding to explicitly model the spatial footprint of glossy BSDF lobes, thereby capturing reflection continuity across materials with varying glossiness, from highly polished to low-sheen surfaces. Unlike previous approaches that treat a ray–surface intersection as an infinitesimal point, NCR considers the reflected ray cone on the scene surface through continuous spatial aggregation. %whose size is determined jointly by the surface roughness and the travel distance of the reflected ray.
To avoid the discontinuities and hard edges that arise from single-point sampling, we introduce a multi-resolution hash grid network \cite{hu2023tri} that takes the footprint’s center and spatial extent as inputs to approximate the pre-filtered radiance distribution.

Glossy reflections often accumulate contributions from a wide range of depths and surfaces, which cannot be accurately represented by a single projected point. To better approximate the BRDF lobe’s integral, NCR traces multiple rays within the cone and clusters the resulting surface hits into several representative groups. The radiance from each cluster is then averaged with appropriate weights to produce the final glossy contribution.

By offloading the burden of modeling complex directional distributions to this pre-filtered module, NCR preserves the fine detail and continuity of reflections for surfaces with diverse glossiness, while also reducing the size of the primary NR network. As a result, it achieves significantly higher visual fidelity on glossy surfaces at a runtime comparable to vanilla NR, offering an efficient and general solution for real-time neural global illumination across a wide range of glossy materials.

Overall, our main technical contributions include:

\begin{itemize}
\item We propose a pre-filtered radiance model that extends neural radiosity to handle glossy materials more efficiently.
\item We introduce a clustering-based approximation for integrating reflected contributions across the glossy BSDF lobe.
\item We present a network architecture that integrates both primary surface interactions and the projected contributions of glossy reflections.
\end{itemize}

\section{Related Works}

\subsection{Neural Scene Representation}

%Neural networks have been widely adopted in graphics community to represent 3D scenes, including Neural Radiance Fields (NeRF) for novel-view synthesis \cite{mildenhall2021nerf}, rendering \cite{muller2021real}, and geometry optimization \cite{muller2022instant}. The most common neural scene representation is coordinate-based neural network \cite{tancik2020fourier}, which takes an coordinate as input, and predict the value on this point. Coordinate-based neural network could represent complex 2D or 3D distributions with high frequency details. Many of their variants intended to improve their performance or extend their capabilities to accommodate more applications.

Neural networks have been widely adopted in the computer graphics community for 3D scene representation, including Neural Radiance Fields (NeRF) for novel view synthesis~\cite{mildenhall2021nerf}, real-time rendering~\cite{muller2021real}, and geometry optimization~\cite{muller2022instant}. The most common formulation is the \emph{coordinate-based neural network}~\cite{tancik2020fourier}, which takes a spatial coordinate as input and predicts the corresponding scene value. Such networks are capable of representing complex 2D and 3D functions with high-frequency details, and have been extended in many ways to improve efficiency and expand applicability.

% To improve rendering speed, early works substitute the large MLP used in NeRF with much smaller ones, and utilized spatial data structures to store local neural features \cite{liu2020neural, reiser2021kilonerf, yu2021plenoctrees}. However, these methods still require training a complete model with large MLP and distilling, resulting in even longer training time and large memory overhead. Later, some works attempted to bypass the original model by directly optimizing local feature with tri-linear interpolation on a voxel grid \cite{fridovich2022plenoxels, sun2022direct}. To reduce large memory budget caused by 3D data structure, sparse data structures have been adopted since only the data on a 2D manifold is required to be accurately modeled in most cases. Some works utilized tensor decomposition represent sparse 3D data in 2D feature planes \cite{chen2022tensorf, tang2022compressible}. This method could be further extended into higher dimension, in order to represent dynamic scenes \cite{fridovich2023k, su2024dynamic}. In 2022, Mueller et al. proposed Instant-NGP \cite{muller2022instant}, which utilized multi-resolution hash grid as scene representation. With these improvements, the time consumption of fitting a 3D scene with neural representation reduced from several hours into minutes or even seconds. The rendering time for a single image has reduced from tens of seconds into less than 0.1 second, enabling its application in interactive tasks.

To accelerate training and inference, early methods replaced large MLPs in NeRF with smaller networks and employed spatial data structures to cache local features~\cite{liu2020neural, reiser2021kilonerf, yu2021plenoctrees}. However, these methods still required distillation from a fully-trained NeRF, resulting in longer training times and high memory usage. Subsequent approaches bypassed pretraining altogether by directly optimizing trilinear features on voxel grids~\cite{fridovich2022plenoxels, sun2022direct}. To reduce memory consumption inherent to 3D structures, sparse representations were proposed, as most visual content lies on a 2D surface. Some works used tensor decompositions to compress 3D volumes into multiple 2D planes~\cite{chen2022tensorf, tang2022compressible}, which could be further extended to higher-dimensional spaces for dynamic scene modeling~\cite{fridovich2023k, su2024dynamic}. A major breakthrough came with Instant-NGP~\cite{muller2022instant}, which utilized a multi-resolution has grid to encode scene features. This drastically reduced the training time from hours to minutes and enabled image rendering within 0.1 seconds, facilitating interactive applications.

% Apart from these dedications in improving performance, many other works extend the model to fit in more tasks. Barron et al. proposed anti-aliased neural radiance field \cite{barron2021mip} with cone tracing. This greatly improved the reconstruction quality of neural representation without additional overhead. Follow up works extended cone tracing for grid-based models, enabling better performance with better quality \cite{hu2023tri, barron2023zip}. Prior neural scene representations only encode directional input with few parameters. Although this was claimed as a prior for 3D view-consistency, it leads to poor reconstruction quality in objects with reflection. Verbin et al. mitigated this issue by explicitly modeling surface normal and roughness with more sophisticated input encoding \cite{verbin2022ref}. Guo et al. used separate models to reconstruct objects in planer reflection areas \cite{guo2022nerfren}. These methods inspired us to customize coordinate-based neural network to fit better into the rendering task.

Beyond speed and memory optimization, several works sought to extend neural representations to better handle aliasing and directional effects. Mip-NeRF~\cite{barron2021mip} introduced cone tracing to model anti-aliasing, while follow-up works extended these ideas to grid-based models for improved rendering quality~\cite{hu2023tri, barron2023zip}. However, these models often use very low-dimensional encodings for directional input. Although this serves as a smoothness prior and improves view consistency, it limits the ability to model sharp directional variations such as specular reflections. To address this, Verbin et al.~\cite{verbin2022ref} introduced explicit modeling of surface normals and roughness, and Guo et al.~\cite{guo2022nerfren} proposed separate models for planar reflection regions.

These limitations motivate our work: we extend coordinate-based neural representations to more faithfully model high-frequency directional effects, particularly those from glossy reflections, while maintaining compatibility with efficient grid-based encodings.

\subsection{Neural Rendering}

% The application of neural networks in rendering work mainly focuses on several aspects: enhancing the results of traditional rendering methods, directly predicting rendering results, and establishing high-dimensional cache structures.

Neural rendering methods typically fall into three categories: post-processing traditional rendering outputs, directly predicting final rendered results, and constructing high-dimensional radiance caches.

% Due to its capability in processing images, early works have utilized convolution neural networks to conduct post-processing on rendered results, such as denoising images from Monte-Carlo path tracer \cite{huo2021survey}, or up sampling for low resolution results \cite{wu2023extrass, zhong2023fusesr}. These methods work by either directly taking the raw images as input and giving results \cite{icsik2021interactive}, or providing a local operator (kernel) for images under certain data distribution \cite{vogels2018denoising}. Other works directly predict rendered images with G-Buffers and direct illumination result as input \cite{nalbach2017deep, granskog2020compositional, diolatzis2022active}.

Early works applied convolutional neural networks (CNNs) to denoise Monte Carlo rendering outputs~\cite{huo2021survey}, perform super-resolution~\cite{wu2023extrass, zhong2023fusesr}, or learn data-dependent filters for local shading refinement~\cite{vogels2018denoising}. These methods either take the rendered images directly as input~\cite{icsik2021interactive}, or operate on auxiliary buffers such as G-buffers and direct lighting~\cite{nalbach2017deep, granskog2020compositional, diolatzis2022active}.

% In addition, the aforementioned coordinate-based neural network has also been widely used in rendering topic. Muller et al. proposed using a neural network to predict the integrand for control variate, and a neural importance sampler \cite{muller2019neural} to estimate the residual between predicted results and ground truth \cite{muller2020neural}. Later, this model served as a neural radiance cache \cite{muller2021real, muller2022instant} for real-time path tracing. This model could be trained online and evaluated in real-time. Dong et al. proposed using similar network to predict a parametric mixture for path guiding \cite{dong2023neural, dong2024efficient}. Hadadan et al. extended the radiosity algorithm \cite{goral1984modeling} into neural radiosity, which utilized the rendering equation as supervision of the output radiance. Follow up works enabled modeling radiance distribution in dynamic scenes with higher dimensional hash grids \cite{coomans2024real}, multiple feature planes with fourier encoding \cite{su2024dynamic}, and deformable feature planes \cite{zheng2024neural}.

Coordinate-based networks have also gained traction for direct radiance prediction. Müller et al.~\cite{muller2019neural} proposed learning control variates and neural importance samplers for path tracing~\cite{muller2020neural}, which later formed the basis for neural radiance caches~\cite{muller2021real, muller2022instant} capable of online training and real-time evaluation. Dong et al.~\cite{dong2023neural, dong2024efficient} used similar architectures to learn mixture models for path guiding. Hadadan et al. introduced Neural Radiosity~\cite{hadadan2021neural}, which reformulates the classical radiosity algorithm~\cite{goral1984modeling} using neural networks and enforces physical correctness via the rendering equation. Subsequent works extended this to dynamic scenes via higher-dimensional hash grids~\cite{coomans2024real}, multiple feature planes with Fourier encoding~\cite{su2024dynamic}, and deformable latent grids~\cite{zheng2024neural}.

% Recently, generative models are utilized to predict radiance distribution under arbitrary scene layout. Zheng et al. decompose overall radiance into background radiance and interactions between different objects \cite{zheng2023nelt}. This method enable high-quality rendering under dynamic scene layout, but it still requires days of training for one single scene. Zeng et al. utilized transformer architecture and interpret triangle features into tokens, enabling rendering arbitrary new objects \cite{zeng2025renderformer}. Although this method requires training on a massive dataset and significant limitations exist on the scene layout, it provides a new potential solution for global illumination.

Recently, generative models have also been explored for radiance prediction under novel scene layouts. Zheng et al.~\cite{zheng2023nelt} decomposed total radiance into background and inter-object interaction terms, while Zeng et al.~\cite{zeng2025renderformer} proposed a Transformer-based architecture that treats triangle features as tokens for scene-agnostic rendering. Despite impressive results, these methods are still too computationally expensive for interactive applications.

Our method positions itself between traditional path tracing and generative models: we adopt an efficient neural architecture for radiance prediction and extend neural radiosity with directional filtering to more accurately represent glossy inter-reflections, while preserving interactive rendering performance.

\subsection{Global Illumination for Glossy Material}

Real-time global illumination (GI) for glossy surfaces remains a challenging task due to the view-dependent nature of specular reflection. Some early methods addressed this using precomputed radiance transfer for distant illumination with single-bounce reflections~\cite{sloan2002precomputed, xu2014practical}. For multi-bounce glossy effects, low-sample path tracing followed by denoising has been widely used~\cite{schied2017spatiotemporal}. However, denoising-based approaches often suffer from temporal instability and flickering artifacts in interactive settings.

More recent works explore combining real-time reflection search with light probe techniques~\cite{guo2022efficient}, yet they still rely on post-process neural denoisers. While these techniques work well in many practical applications, they either struggle with secondary bounce accuracy or incur additional latency due to multi-pass filtering.

Existing neural GI methods towards glossy materials can be broadly divided into three streams. Approaches such as Neural Radiance Caching (NRC)~\cite{muller2021real, muller2022instant} and the RHS formulation of Neural Radiosity~\cite{hadadan2021neural} only query the network at secondary hit of rays. In this setting, glossy effects do not require special treatment, but the supervision is indirect, and the final renderings often exhibit noticeable noise. By contrast, LHS variant of NR evaluates the network directly at primary intersections, yet relies on simple directional encodings to fit the outgoing radiance distribution. As a result, highly glossy, view-dependent effects remain difficult to reconstruct. More recent extensions, such as NeLT~\cite{zheng2023nelt} and LightFormer~\cite{ren2024lightformer}, enrich the NR framework with explicit encodings of light source positions, enabling them to reproduce sharp highlights caused by direct lighting. However, they still struggle with high-frequency glossy features that arise from multi-bounce indirect illumination, where the reflected lobes are more complex and spatially varying.

In contrast, our method addresses the challenge of glossy GI by incorporating \emph{ray cone encoding} directly into the neural radiosity framework. This allows us to prefilter radiance according to surface roughness and reflection geometry, leading to better reconstruction of high-frequency directional effects such as caustics and sharp inter-reflections, all within a lightweight and efficient network design.

\section{Preliminary}

\subsection{Rendering Equation}

Physics-based rendering is fundamentally governed by the rendering equation \cite{kajiya1986rendering}, which expresses the outgoing radiance as a combination of self-emission and the integral of incident radiance over the hemisphere:
\begin{equation}
L_o(\mathbf{x},\omega_o) = L_e(\mathbf{x},\omega_o) + \int_{\mathcal{H}^2} L_i(\mathbf{x},\omega_i) f_s(\mathbf{x},\omega_i,\omega_o) \lvert\mathbf{n}\cdot\omega_i\lvert \mathrm{d}\omega_i,
\end{equation}

\noindent
where $L_o$, $L_i$, and $L_e$ denote outgoing radiance, incident radiance, and self-emission, respectively. The pair $(\mathbf{x},\omega_o)$ or $(\mathbf{x},\omega_i)$ represents the radiance at surface point $\mathbf{x}$ in the outgoing or incident direction $\omega_o$ or $\omega_i$. The integral accumulates the contribution of incident light scattered toward the outgoing direction across the hemisphere $\mathcal{H}^2$. The bidirectional scattering distribution function (BSDF) $f_s$ characterizes how light is reflected from direction $\omega_i$ to $\omega_o$ at point $\mathbf{x}$.

Due to the linear nature of light transport, outgoing radiance at one surface point can serve as incident radiance at another. This property makes the rendering equation inherently recursive. Solving this recursive equation with a spherical integral is computationally expensive and forms the core challenge of physically-based rendering.

\subsection{Neural Radiosity}

To address this challenge, NR \cite{kajiya1986rendering} models the outgoing radiance distribution using a neural network, and leverages the rendering equation itself as supervision during training. Specifically, the network’s prediction aims to minimize the residual between the left-hand side (LHS) and right-hand side (RHS) of the rendering equation:
\begin{equation}
L_{LHS}(\mathbf{x},\omega_o;\Theta) = L_\Theta(\mathbf{x},\omega_o),
\end{equation}
\begin{multline}
L_{RHS}(\mathbf{x},\omega_o;\Theta) = L_e(\mathbf{x},\omega_o) \\
+ \int_{\mathcal{H}^2} L_\Theta(\mathbf{x}'(\mathbf{x},\omega_i), -\omega_i)
    f_s(\mathbf{x},\omega_i,\omega_o) \lvert\mathbf{n}\cdot\omega_i\lvert \mathrm{d}\omega_i \ ,
\end{multline}
\begin{equation}
r_\Theta(\mathbf{x},\omega_o) = L_{LHS}(\mathbf{x},\omega_o;\Theta) - L_{RHS}(\mathbf{x},\omega_o;\Theta) \ ,
\end{equation}
\noindent
where $\Theta$ represents the parameters of the neural network, and $\mathbf{x}'(\mathbf{x},\omega_i)$ denotes the intersection point of an incident ray originating from $\mathbf{x}$ in direction $\omega_i$ with the nearest surface. The RHS integral is approximated via Monte Carlo integration using $M$ incident ray samples:
\begin{multline}
L_{RHS}(\mathbf{x},\omega_o;\Theta) \approx L_e(\mathbf{x},\omega_o) + \\
 \frac{1}{M} \sum_{m=1}^M L_\Theta(\mathbf{x}'(\mathbf{x},\omega_{i,m}), -\omega_{i,m})
    f_s(\mathbf{x},\omega_{i,m},\omega_o) \lvert\mathbf{n}\cdot\omega_{i,m}\lvert.
\end{multline}

The network is optimized using a relative mean squared error (rMSE) loss function defined as:
\begin{equation}
m_\Theta(\mathbf{x},\omega_o) = \frac{L_{LHS}(\mathbf{x},\omega_o;\Theta) + L_{RHS}(\mathbf{x},\omega_o;\Theta)}{2},
\end{equation}
\begin{equation}
\mathcal{L}_{rMSE} = \frac{1}{N}\sum_{j=1}^{N} \lVert\frac{r_\Theta(\mathbf{x}_j,\omega_{o,j})}{\mathrm{sg}(m_\Theta(\mathbf{x}_j,\omega_{o,j})) + \epsilon}\rVert^2 \ ,
\label{eq:nr}
\end{equation}

\noindent
where $\mathrm{sg}(\cdot)$ is the stop-gradient operator, which prevents the denominator from influencing the gradient flow, thereby stabilizing training.

The architecture of NR comprises two primary components: a multi-resolution feature encoding and a compact multi-layer perceptron (MLP). The feature encoding consists of $L$ 3D grids at increasing resolutions, which take the position $\mathbf{x}$ as input. To evaluate the feature at a given location $x \in \mathbb{R}^3$, the values at the eight corners of the enclosing voxel are tri-linearly interpolated. Features from all resolutions are concatenated into a single vector $\mathbf{v}(\mathbf{x})$. This vector, along with other attributes such as the outgoing direction $\omega_o$, surface normal $\mathbf{n}$, material reflectance (diffuse or specular) $\alpha$, and roughness $\rho$, is then passed to the MLP for radiance prediction.

By incorporating direction and surface properties into the input, NR can handle a variety of materials, including microfacet-based ones. However, due to the spectral bias~\cite{rahaman2019spectral} inherent in neural networks, NR struggles to reproduce high-frequency angular detail, an essential requirement for accurately rendering glossy materials such as polished metals, plastics, and varnished wood. To overcome this limitation, our work extends NR with ray cone encoding and introduces a corresponding approximation algorithm to improve the fidelity of directional reflectance modeling.

\subsection{Dilemma for Coordinate-based Neural Networks}

We aim to optimize neural scene representations for interactive global illumination. Coordinate-based neural networks are widely adopted in neural rendering and novel view synthesis due to their ability to represent high-frequency details in low-dimensional spaces \cite{tancik2020fourier} (typically 2D or 3D). Moreover, their compactness allows efficient parallel inference across large numbers of pixels.

Despite these advantages, coordinate-based neural networks face challenges in modeling directional distributions accurately. A key reason is that directional inputs are typically encoded into much lower-dimensional representations than positional inputs. On the one hand, this design choice aligns with the prior that outgoing radiance distributions are generally smooth and low-frequency, particularly for Lambertian surfaces. This assumption works well for radiance caching methods that query the network indirectly, such as the RHS formulation of NR or NRC.

However, evaluating the network at secondary intersections introduces sampling noise into the rendered image. Reducing this noise requires either increasing the number of samples per pixel or applying post-processing denoising techniques, both of which significantly increase computational cost. In interactive rendering scenarios, this often leads to undesirable trade-offs, such as high latency or visible flickering artifacts.

\section{Method}

%In this section, we first introduce a cone-encoding strategy to model the view-dependent nature of glossy reflections, which are difficult to capture with traditional position-based encodings. Next, we propose a clustering-based approximation, since the cone–surface intersection induced by cone encoding lacks an analytic form and cannot be computed directly. Finally, we present the network architecture that builds upon this representation.

In this section, we introduce the mechanism of Neural Cone Radiosity (NCR). NCR models outgoing radiance with neural networks and captures view-dependent effects through a reflection-aware cone encoding. We first formulate the cone during tracing, as long as the corresponding encoding for glossy surface interactions (Sec. \ref{sec:coneencoding}). Then, a clustering-based approximation is used to decompose the cone’s projection on the scene into multiple regions, each of which can be efficiently represented using feature-grid neural networks (Sec. \ref{sec:clustering}). Once combined, these features provide strong representational power for modeling the outgoing radiance distribution of glossy reflections. To achieve both efficiency and adaptability in representing different materials, we finally introduce a dual-branch radiance model that separates diffuse and glossy components, with a lightweight modulation network blending them to handle arbitrary surface roughness (Sec. \ref{sec:network}). 

\begin{figure}
    \centering
    \includegraphics[width=\linewidth]{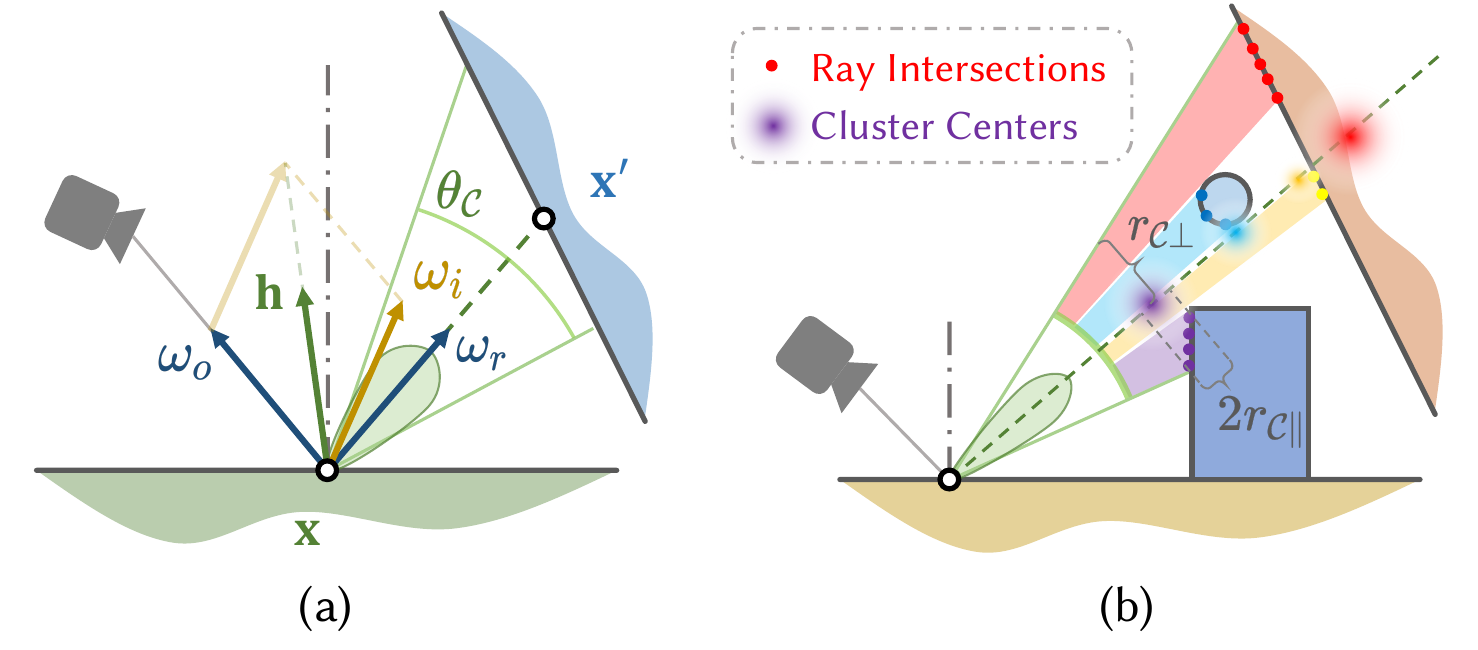}
    \caption{Illustration of cone encoding on a glossy surface. (a) Glossy reflection. The glossy BRDF lobe can be bounded by a cone centered on the reflection direction, with its aperture determined by surface roughness. (b) We illustrate the clustering approximation mechanism: the intersections of reflected rays with the scene surface are depicted with small dots, and their marching distances are aggregated to those of the cluster centers. For simplicity, each cluster center is assumed to lie along the specular reflection direction.}
    \label{fig:cone}
\end{figure}

\subsection{Cone Encoding for Glossy Surfaces}
\label{sec:coneencoding}
Cone encoding is the key to model view-dependent glossy reflections as prefiltered incident radiance over the BSDF lobe.
Our cone encoding elevates radiance sampling from a local, point-wise evaluation to a scale-aware spatial integration process. Instead of treating reflection as a single interaction, it models the finite spatial footprint induced by the reflection cone, thereby embedding a continuous aggregation of radiance across this region. This shift transforms point sampling into a principled operator that links angular reflectance to its spatial manifestation, providing an intrinsic scale-adaptive filtering mechanism. Conceptually, cone encoding unifies geometry, reflection lobes, and radiance sampling under a single framework for theoretically coherent representation of glossy transport (see Figure \ref{fig:cone}).

From the rendering equation, we observe that the integral term can be interpreted as a convolution between the incident radiance and the BSDF at point $\mathbf{x}$. While glossy surfaces exhibit more complex outgoing radiance distributions compared with diffuse surfaces, their BSDFs have narrower support. This implies that the reflected radiance can be approximated by pre-filtering the incident radiance over the BSDF's support region, which is centered around the specular reflection direction.

To account for the spatial extent of the reflected lobe, we trace a cone $\mathcal{C}(\mathbf{x}, \omega_r, \theta_\mathcal{C})$ at each surface interaction instead of a single ray. Here, $\omega_r = 2\lvert\mathbf{n}\cdot\omega_o\rvert\mathbf{n} - \omega_o$ represents the specular reflection direction, and the cone angle $\theta_\mathcal{C} = 2\langle\omega_r, \omega_\tau\rangle$ is determined by the surface roughness $\rho$ and the normal distribution function (NDF) of the microfacet BSDF model at point $\mathbf{x}$.

In our formulation, the cone is defined as the level set of the NDF $D(\theta, \rho)$ such that the integral of the NDF within the cone reaches a predefined threshold $\tau$ (see Figure~\ref{fig:cone}a):
\begin{equation}
\int_0^{2\pi}\int_0^{\theta_\mathcal{C}}D(\theta, \rho)\mathrm{d}\theta\mathrm{d}\phi = \tau \int_0^{2\pi}\int_0^{\frac{\pi}{2}}D(\theta, \rho)\mathrm{d}\theta\mathrm{d}\phi,
\label{eq:ndf}
\end{equation}

\noindent
where $\theta = \langle\mathbf{n}, \mathbf{h}\rangle$ is the angle between the surface normal $\mathbf{n}$ and the half-vector $\mathbf{h} = \frac{\omega_i + \omega_o}{\lVert \omega_i + \omega_o \rVert}$, and $\phi$ denotes the azimuthal angle.

%Since the relationship between $\rho$ and $\theta_\mathcal{C}$ does not admit an analytical solution, we precompute the threshold angles $\theta_\mathcal{C}$ via numerical integration for roughness values $\rho$ ranging from 0 to 0.5, and store them in a lookup table. During rendering, the angle values are retrieved by interpolating from this table.

Since the relationship between $\rho$ and $\theta_C$ does not admit a closed-form expression, 
we approximate it numerically. Specifically, we compute $\theta_C(\rho)$ for $\rho \in [0,0.5]$ 
via numerical integration, and represent the result as a discretized functional mapping 
$\mathcal{D} : [0,0.5] \to \mathbb{R}$, where $\mathcal{D}(\rho_i) = \theta_C(\rho_i)$ at sampled points $\{\rho_i \}$.
During rendering, $\theta_C(\rho)$ is evaluated by interpolating within $\mathcal{D}$, 
yielding a continuous approximation of the underlying function.

Building on this cone formulation, we generalize the conventional point-sampling operator into a cone-aware query that explicitly accounts for the finite spatial support of the reflected radiance lobe. Instead of evaluating features at the primary surface intersection $\mathbf{x}$, we define the query domain as the projected intersection of the reflection cone with the scene surface, centered at the secondary hit point $\mathbf{x}' = \mathbf{x} + t \omega_r$, where $t$ denotes the distance along the reflection direction $\omega_r$. 
The induced footprint is modeled as a disk of radius
$r_\mathcal{C\perp}=t\cdot {\tan(\frac{\theta_\mathcal{C}}{2})}$, 
representing the orthogonal projection of the cone aperture.

%The projected cone radius $r_{\mathcal{C}\perp}$ is proportional to the distance  between $\mathbf{x}$ and $\mathbf{x}'$:

Given the footprint parameters ($\mathbf{x}'$, $r_\mathcal{C\perp}$), the radiance representation is then queried over this spatial support, yielding a scale-aware encoding of glossy transport. We denote this operator as the cone encoding, which extends point-sampling to a continuous spatial aggregation consistent with the geometry of glossy reflection.

%Inspired by previous work about novel view synthesis that aimed at anti-aliasing \cite{}, we adopt a multi-resolution feature grid to model the reflected radiance. Unlike previous methods that reduce feature vectors from every resolution level (mean-reduce or concatenation), we interpolate feature vector $\mathbf{v}_{glo}(\mathbf{x}, r_\mathcal{C})$ from the two levels with adjacent resolutions with respect to the queried scale $r_\mathcal{C}$:
Inspired by anti-aliasing strategies in novel view synthesis \cite{barron2021mip, hu2023tri}, we adopt a multi-resolution feature grid to represent reflected radiance. Unlike prior methods that aggregate feature vectors from all resolution levels via mean-reduction or concatenation, we interpolate the feature vector $\mathbf{v}_{glo}(\mathbf{x}, r_\mathcal{C})$ from the two grid levels with resolutions closest to the queried scale $r_\mathcal{C}$:
\begin{equation}
\mathbf{v}_{glo}(\mathbf{x},r_\mathcal{C}) = 
\frac{s\cdot r_\mathcal{C} - l_{i+1}}{l_{i} - l_{i+1}} \mathbf{v}_i(\mathbf{x}) +
\frac{l_i - s\cdot r_\mathcal{C}}{l_{i} - l_{i+1}} \mathbf{v}_{i+1}(\mathbf{x}),
\label{eq:v_glo}
\end{equation}
\begin{equation}
r_\mathcal{C} = r_\mathcal{C\perp} + r_\mathcal{C\parallel},
\label{eq:rc}
\end{equation}

\noindent
%where $l_i$ and $l_{i+1}$ are the grid resolutions at levels $i$ and $i+1$, satisfying $l_{i+1} \leq s \cdot r_\mathcal{C} \leq l_i$. The feature vector $\mathbf{v}_i(\mathbf{x})$ is queried at level $i$ using tri-linear interpolation. The queried scale $r_\mathcal{C}$ is defined as the mean of the projected cone radius $r_{\mathcal{C}\perp}$ and the axial scale $r_{\mathcal{C}\parallel}$, which will be discussed in the next subsection. The sampling ratio $s$ is a hyperparameter that defines the mapping between the grid resolution and the filter size. By default, it is set to 2, which means the grid is slightly larger than the filter. This mechanism serves as a smooth prior for glossy model. For scenes where the cone–surface intersection area is excessively large or small, $s$ can be adjusted to ensure that sample points are more evenly distributed across all resolution layers.
where $l_i$ and $l_{i+1}$ are the grid resolutions at levels $i$ and $i+1$, respectively, satisfying the condition $l_{i+1} \leq s \cdot r_\mathcal{C} \leq l_i$. The feature vector $\mathbf{v}_i(\mathbf{x})$ is queried at level $i$ via trilinear interpolation. The queried scale $r_\mathcal{C}$ is defined as the sum of the projected cone radius $r_{\mathcal{C}\perp}$ and the axial scale $r_{\mathcal{C}\parallel}$, which will be described in the next subsection.

The sampling ratio $s$ is a hyperparameter that defines the mapping between the grid resolution and the filter size. By default, it is set to 1, meaning that the grid is about the same size as the filter. This design acts as a smoothness prior for the glossy model. For scenes where the cone–surface intersection area is unusually large or small, $s$ can be adjusted to ensure that sample points are more evenly distributed across all resolution levels.

\subsection{Clustering Approximation}
\label{sec:clustering}
In general, the intersection between the reflected cone and scene geometry does not result in a perfect circular footprint due to non-perpendicular surface angles and complex geometry (see Figure~\ref{fig:cone}b). Analytically computing the exact domain of this intersection is computationally intractable. Therefore, we approximate the footprint by dividing it into multiple subdomains, each of which can be locally represented as a spherical proxy that can be efficiently modeled using the cone encoding operator described earlier.

%To achieve this, we trace $T$ reflected rays with importance sampling on the glossy BSDF lobe, and divide them into $K$ clusters with K-Means algorithm \cite{}. Then we query the reflection network with these clusters to get their reflected radiance $L_r$, and use weighted radiance as the output of the glossy model, where the weight is the proportion of rays in this cluster $T_k$ to the total number of rays:
To achieve this, we trace $T$ reflected rays from the surface point using importance sampling over the glossy BSDF lobe, thereby capturing the stochastic support of the reflection distribution. These resulting rays are then grouped into $K$ clusters using the K-Means algorithm~\cite{ahmed2020k}, yielding representative spatial supports. For each cluster, we query the reflection network to obtain the reflected radiance $L_r$, and compute the final output of the glossy reflection model as the weighted sum of these radiance values. The weight assigned to each cluster corresponds to the proportion of rays $T_k$ in that cluster relative to the total number of rays:
\begin{equation}
L_{glo}(\mathbf{x}, \omega_o) = \sum_{k=1}^K \frac{T_k}{T} L_r(\mathbf{x}'_k, -\omega_r, r_{\mathcal{C},k}),
\end{equation}

\noindent
%By leveraging cluster weights, we can obtain a smooth transition between two clusters with significant differences in radiance. This has a great advantage in rendering soft shadows and when there is a major depth change in incident distribution.
where $k$ is the index of each cluster. Equation~\ref{eq:v_glo},\ref{eq:rc} are applied to each cluster to get their corresponding radius $r_{\mathcal{C},k}$. By applying these cluster-based weights, we achieve smooth transitions across clusters, even when there are large radiance differences. This formulation is particularly advantageous in rendering soft shadows and scenarios with significant depth variation in the incident distribution.

%To reduce computational overhead, we conduct 1D K-Means for the reflected rays on their marching distance $t_t$ instead of 3D K-Means on their position coordinate. Cluster centers are decided by their average marching distance $t_k$:
To reduce computational overhead, we perform one-dimensional K-Means clustering based on the marching distances $t_t$ of the reflected rays, instead of conducting three-dimensional clustering on their intersection coordinates. The center of each cluster is computed as the mean marching distance $t_k$:
\begin{equation}
\mathbf{x}'_k = \mathbf{x} + t_k\omega_r,\
\mathrm{where}\
t_k = \frac{1}{T_k} \sum_{t=1}^{T_k}t_t \ .
\end{equation}

The axial scale $r_{\mathcal{C\parallel},k}$ for each cluster is estimated as the standard deviation of the forward distances:
\begin{equation}
r_{\mathcal{C\parallel},k} = \sqrt{\frac{1}{T_k}\sum_{t=1}^{T_k}t_t^2 - t_k^2} \ .
\end{equation}

\subsection{Network Architecture}
\label{sec:network}
\begin{figure}
    \centering
    \includegraphics[width=\linewidth]{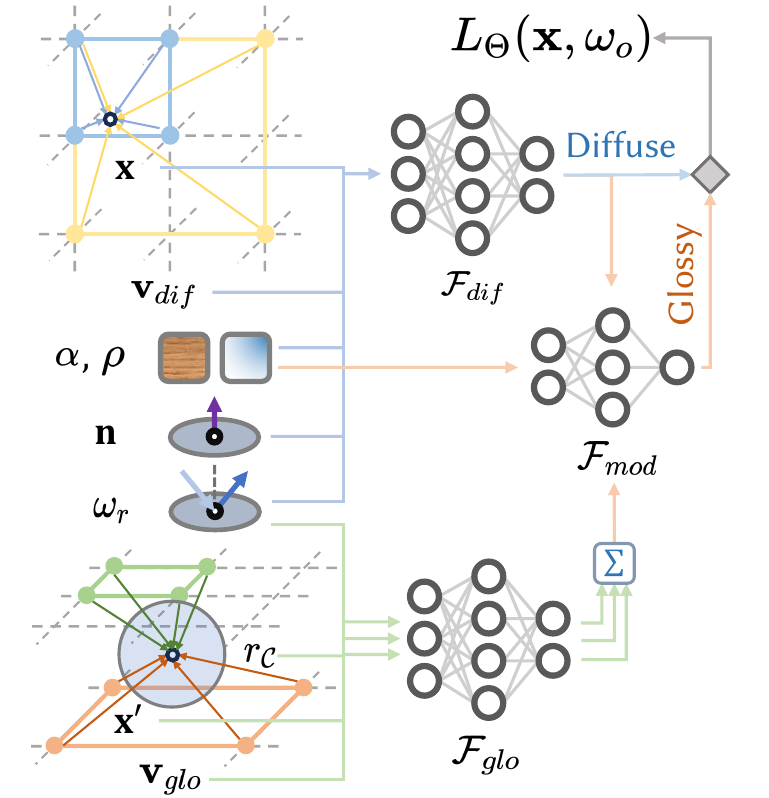}
    \caption{Our network architecture. The system comprises a diffuse network $\mathcal{F}_{dif}$, a glossy network $\mathcal{F}_{glo}$, and a modulation network $\mathcal{F}_{mod}$. $\mathcal{F}_{dif}$ takes multi-resolution hash grid features at position $\mathbf{x}$, along with surface normal $\mathbf{n}$, roughness $\rho$, reflectance $\alpha$, and specular direction $\omega_r$. For each cluster center $\mathbf{x}'$, features are interpolated into $\mathbf{v}_{glo}$ and passed to $\mathcal{F}_{glo}$ along with $\mathbf{x}'$, cluster size $r_\mathcal{C}$, and $\omega_r$ to predict per-cluster output, which are aggregated into a glossy prediction. For glossy surfaces, $\mathcal{F}_{mod}$ combines diffuse and glossy predictions based on $\rho$ and $\alpha$, while for diffuse surfaces, the diffuse prediction is used directly.}
    \label{fig:network}
\end{figure}

We adopt a dual-branch architecture. One branch (similar to NR) predicts diffuse outgoing radiance, and the other, a more compact branch, models glossy reflections. To handle arbitrary surface roughness, a lightweight modulation network $\mathcal{F}_{mod}$ is introduced to blend the outputs from the diffuse and glossy branches (See Figure~\ref{fig:network}).

%The diffuse neural network is composed of a multi-resolution hash grid encoding $\mathbf{v}_{dif}(\mathbf{x})$, and a small MLP $\mathcal{F}_{dif}$. Each layer of the hash grid produces a feature vector tri-linearly interpolated from the vertices of voxel grid it resides. All the vectors from each layer are concatenated into a long vector $\mathbf{v}_{dif}(\mathbf{x})$. Finally, this vector along with the position $\mathbf{x}$, outgoing direction $\omega_o$, surface normal $\mathbf{n}$, roughness $\rho$, and material reflectance (diffuse or specular) $\alpha$, is then passed to the MLP $\mathcal{F}_{dif}$ for diffuse radiance prediction:
As diffuse radiance mainly varies with spatial position, we follow previous practices~\cite{hadadan2021neural, muller2022instant} and adopt a feature-grid-based network. The diffuse network consists of a multi-resolution hash grid encoder $\mathbf{v}_{dif}(\mathbf{x})$ and a small MLP $\mathcal{F}_{dif}$. Each level of the hash grid produces a feature vector via trilinear interpolation of the eight surrounding voxel vertices. All level-wise features are concatenated into a single vector $\mathbf{v}_{dif}(\mathbf{x})\in \mathbb{R}^{d\times l}$. This vector, along with auxiliary inputs including position $\mathbf{x}\in \mathbb{R}^{3}$, outgoing direction $\omega_o\in [-1,1]^{3}$, surface normal $\mathbf{n}\in [-1,1]^{3}$, roughness $\rho\in \mathbb{R}$, and reflectance coefficient $\alpha\in [0,1]^{3}$, is fed into the diffuse MLP:
\begin{equation}
L_{dif}(\mathbf{x}, \omega_o, \mathbf{n}, \rho, \alpha) = \mathcal{F}_{dif}(\mathbf{v}_{dif}(\mathbf{x}), \mathbf{x}, \omega_r, \mathbf{n}, \rho, \alpha),
\end{equation}
\begin{equation}
\mathbf{v}_{dif}(\mathbf{x}) = \bigoplus_{l=1}^L \mathrm{TriLerp}(\mathbf{x}, V_l(\mathbf{x})).
\end{equation}

\noindent
%Note that the directional input here is the specular reflection direction $\omega_r=2\lvert\mathbf{n}\cdot\omega_o\rvert\mathbf{n}-\omega_o$ other than outgoing direction $\omega_o$. This conversion disentangles directional distribution from surface normal, enabling the network to learn the incident radiance distribution convolved by the BSDF lobe, which has better spatial continuity \cite{}.
Note that the actual directional input is the specular reflection direction $\omega_r = 2\lvert\mathbf{n}\cdot\omega_o\rvert\mathbf{n} - \omega_o$, rather than the outgoing direction $\omega_o$. This conversion helps decouple the directional distribution from the surface normal, enabling the network to more effectively learn the incident radiance distribution convolved with the BSDF lobe, which typically exhibits better spatial continuity \cite{verbin2022ref}.

%The glossy neural network also consists of a multi-resolution hash grid and an MLP. As mentioned in Equation \ref{eq:v_glo}, the output of the hash grid is $\mathbf{v}_{glo}(\mathbf{x}, r_\mathcal{C})$. This vector, concatenated along with the cluster center $\mathbf{x}'=\mathbf{x}'_k$, outgoing direction of the secondary intersection surface $\omega'_o=-\omega_r$, and the queried scale $r_\mathcal{C}$ is passed to the glossy MLP $\mathcal{F}_{glo}$ for reflected radiance prediction of one cluster:
Since the queried region for each cluster is scale-dependent, we incorporate the query scale into the input representation to help the network adapt to different surface roughness levels and sampling radii. The glossy network similarly comprises a multi-resolution hash grid and an MLP. As described in Equation~\ref{eq:v_glo}, the hash grid output is $\mathbf{v}_{glo}(\mathbf{x}, r_\mathcal{C})$. This vector, concatenated with the cluster center $\mathbf{x}' = \mathbf{x}'_k$, the outgoing direction of the secondary intersection $\omega'_o = -\omega_r$, and the queried scale $r_{\mathcal{C},k}$, is passed into the glossy MLP $\mathcal{F}_{glo}$ to predict reflected radiance for each cluster:
\begin{equation}
L_{r}(\mathbf{x}'_k, -\omega_r, r_{\mathcal{C},k}) = \mathcal{F}_{glo}(\mathbf{v}_{glo}(\mathbf{x}'_k, r_{\mathcal{C},k}), \mathbf{x}'_k, -\omega_r, r_{\mathcal{C},k}).
\end{equation}

\noindent
%Note that surface normal and reflectance is no more the input of glossy MLP, this is because these variables is under-determined for an area. On the other hand, pre-filtered radiance has smooth spatial distribution, where hash grid is capable enough to learn.
Unlike the diffuse network, the glossy MLP does not take surface normal or material reflectance as input. This is because these quantities are ill-defined over a spatially extended region. Instead, the pre-filtered radiance is assumed to exhibit sufficient spatial smoothness, which the hash grid is capable of modeling effectively.

%The diffuse network offers more accurate prediction when the shaded surface has larger roughness value so that the outgoing radiance distribution is smooth enough for the diffuse MLP to fit. While the glossy network offers more accurate prediction when shaded surface has smaller roughness value, so that the clusters can precisely cover the area the BSDF lobe projects.
The diffuse network performs best when surface roughness is high, resulting in smooth outgoing radiance distributions that the MLP can easily approximate. Conversely, the glossy network excels when surface roughness is low, allowing the cluster-based sampling to precisely capture the projected BSDF lobe.

%Based on this property, we modulate the output from these two networks with another tiny MLP $\mathcal{F}_{mod}$ with roughness $\rho$ and surface reflectance as additional input, to get the final prediction for outgoing radiance:
Based on this complementary behavior, we employ a modulation network $\mathcal{F}_{mod}$—a small MLP that takes as input the outputs from both branches along with roughness $\rho$ and reflectance $\alpha$—to produce the final prediction for outgoing radiance:
\begin{multline}
L_\Theta(\mathbf{x},\omega_o) = \\
\begin{cases}
\mathcal{F}_{mod}(L_{dif}(\mathbf{x}, \omega_o, \mathbf{n}, \rho, \alpha), L_{glo}(\mathbf{x}, \omega_o), \rho, \alpha), & \rho < 0.5 \ , \\
L_{dif}(\mathbf{x}, \omega_o, \mathbf{n}, \rho, \alpha),   &  \rho \geq 0.5 \ .
\end{cases}
\end{multline}

\begin{figure}
    \centering
    \includegraphics[width=\linewidth]{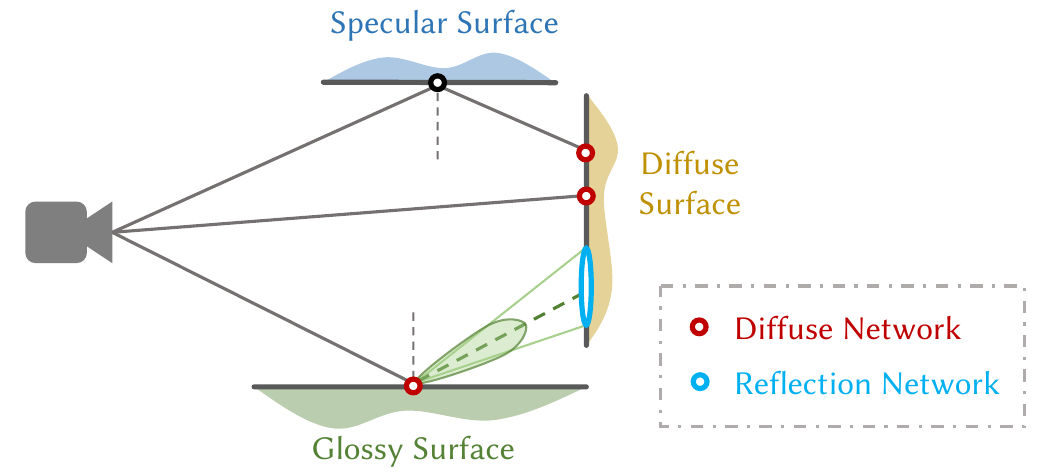}
    \caption{Network inference locations for different surface interactions. All surface interactions are categorized into three types based on surface roughness: specular surfaces ($\rho = 0$), glossy surfaces ($0 < \rho < 0.5$), and diffuse surfaces ($\rho \geq 0.5$). For specular surfaces, secondary rays are traced recursively until a non-specular surface is encountered. For diffuse surfaces, the diffuse network is queried directly to predict radiance. For glossy surfaces, an additional network query is performed using the cone–surface intersection, and its result is combined with the diffuse prediction to produce the final radiance output.}
    \label{fig:path}
\end{figure}

\section{Implementation}

\subsection{Framework and Hardware}

%We implement our work with Mitsuba3 renderer~\cite{jakob2022mitsuba3}. The neural networks are implemented using Pytorch~\cite{paszke2019pytorch} and \texttt{tiny-cuda-nn}~\cite{tiny-cuda-nn}. To gain better rendering performance, we implement the forward pass for multi-resolution feature grid and 1D K-Means algorithm with customized CUDA kernel. We train our model on NVidia RTX 4090 GPU and render images on NVidia RTX 3090 GPU.

Our implementation is based on the Mitsuba 3 renderer~\cite{jakob2022mitsuba3}. Neural network components are implemented using PyTorch~\cite{paszke2019pytorch} and \texttt{tiny-cuda-nn}~\cite{tiny-cuda-nn}. To improve rendering performance, we develop custom CUDA kernels for the forward pass of the multi-resolution feature grid and a 1D K-Means algorithm. Training is conducted on an NVIDIA RTX 4090 GPU, while rendering is performed on an NVIDIA RTX 3090 GPU.

\subsection{Rendering Pipeline}

% To render images with our proposed method, we trace a single ray per pixel towards the scene, to get the geometry information for neural network input. Afterwards, we will use the output image as the texture and draw it on the full screen pass.

To render an image using our method, we trace a single primary ray per pixel towards the pixel centers into the scene to gather geometry information as input to the neural network. The resulting output is then used as a texture in a full-screen pass.

% As in previous work~\cite{hadadan2021neural}, we trace secondary rays for specular surfaces until a non-specular surface is encountered. This practice avoids making MLPs learn complex directional distribution without additional network inference.

Following prior work~\cite{hadadan2021neural}, we trace secondary rays through specular surfaces until a non-specular surface is encountered (see Figure~\ref{fig:path}). This approach avoids forcing the MLP to learn complex directional distributions, which would otherwise require additional network inference.

% Since we only trace a single ray per pixel due to performance concern, noticeable aliasing issue could occur, especially around object silhouettes and texture edges with high contrasts. To minimize aliasing as much as possible, we apply post-process anti-aliasing techniques (FXAA) to the rendered image, and trace all the rays towards the pixel centers. This effectively alleviates aliasing issue, without introducing much overhead.

Due to our single-ray-per-pixel design (for performance reasons), aliasing artifacts may appear—particularly around object silhouettes and regions with high-frequency texture details. To mitigate this, we apply post-process anti-aliasing using FXAA~\cite{fxaa_} and trace rays toward pixel centers. This strategy effectively reduces aliasing while incurring minimal computational overhead.

During training, we randomly sample 65,536 visible surface interactions per batch across the scene, excluding the backsides of objects and regions that are never shaded. For each surface interaction, an outgoing direction is randomly sampled, along with $M = 32$ corresponding incident rays. Following Su et al.~\cite{su2024dynamic}, we progressively double $M$ every quarter of the training schedule while halving the number of surface samples, striking a balance between faster training and stable convergence.

\subsection{Architecture}

%In our experiments, the multi-resolution hash grid of diffuse network $\mathbf{v}_{dif}$ consists of 4 resolution layers, with base resolution of 32 voxels, and the ratio of resolution between two adjacent layers is 2. Its MLP $\mathcal{F}_{dif}$ has 3 hidden layers, each has 128 neurons. The hash grid of glossy network $\mathbf{v}_{glo}$ has 8 layers with a resolution scale of 1.414. Its MLP $\mathcal{F}_{glo}$ has 2 hidden layers, each has 64 neurons. The modulation network is a small network with 1 hidden layer of 32 neurons.

In our experiments, the diffuse network's multi-resolution hash grid $\mathbf{v}_{dif}$ consists of 4 resolution levels, with a base resolution of 32 voxels. Each successive level increases resolution by a factor of 2. The associated MLP $\mathcal{F}_{dif}$ has 3 hidden layers, each with 128 neurons.

The glossy network's hash grid $\mathbf{v}_{glo}$ contains 8 levels with a resolution growth factor of 2 and base resolution of 4. Its MLP $\mathcal{F}_{glo}$ has 2 hidden layers with 64 neurons each. The modulation network is a compact module with a single hidden layer of 32 neurons.

% During training, the number of rays traced for each glossy surface $T$ is 128, and are grouped into 4 clusters to reduce noise. In an interactive renderer, we trace 64 rays and group them into 2 clusters to gain better performance.

During training, we trace $T=128$ rays for each glossy surface, which are grouped into 4 clusters to reduce variance. In the interactive renderer, we reduce this to 32 rays to improve runtime performance. The cone threshold is set to $\tau=0.99$.

% We use ReLU as activation function between hidden layers. Instead of using absolute value as output activation like in NR, we use SquarePlus~\cite{barron2021squareplus}, an smooth alternative for ReLU activation.

We use ReLU as the activation function between hidden layers. For the output layer, instead of using the absolute value activation as in Neural Radiosity, we adopt SquarePlus~\cite{barron2021squareplus}, a smooth and differentiable alternative to ReLU that avoids the discontinuity of hard activations.

\section{Results and Analysis}

\subsection{Comparison}

We conduct comparative experiments on several modified scenes from the Bitterli dataset~\cite{bitterli2016resource}. All images are rendered without any super-resolution post-processing.

\begin{itemize}
\item \emph{Bathroom}: Features a frosted mirror, a shiny tiled floor, and a ceramic bathtub, at 1024$\times$1024 resolution.
\item \emph{Cornell-box}: Features a frosted mirror on the right wall reflecting plastic bunny and dragon models placed on two boxes. The scene includes complex textures on the back wall, left wall, and floor, at 1024$\times$1024 resolution.
\item \emph{kitchen}: Includes a brushed metal extractor hood, ceramic plates, and various complex geometries with glossy surfaces, at 1280$\times$720 resolution.
\item \emph{Living-room}: Contains a variety of varnished wood surfaces and a polished mirror, at 1280$\times$720 resolution.
\item \emph{Veach-ajar}: Comprises a glossy floor and two pots with specular highlights, at 1280$\times$720 resolution.
\end{itemize}

% Quality with NR, PT 4spp + oidn, PT 16spp

\begin{figure*}
    \centering
    \includegraphics[width=\linewidth]{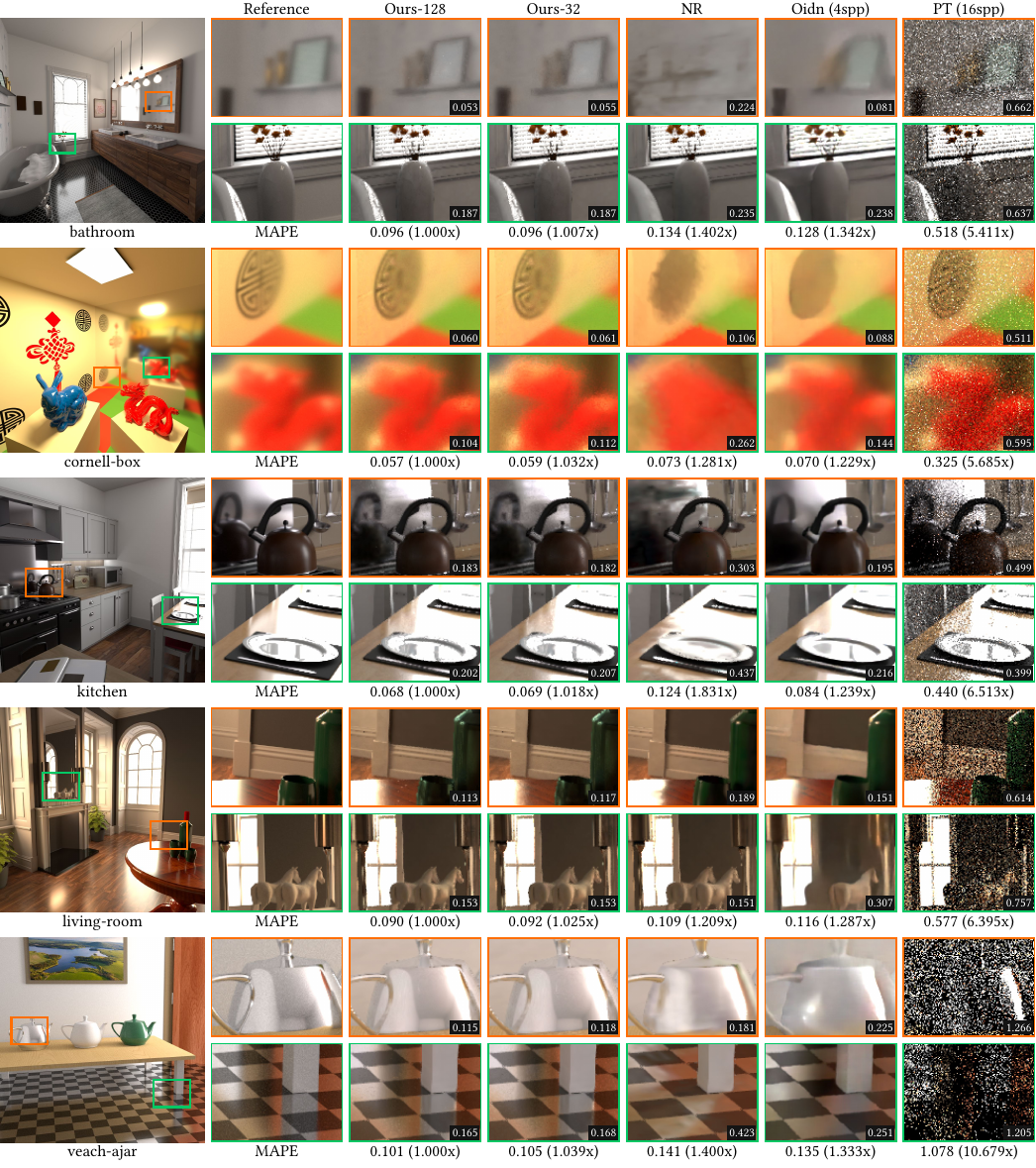}
    \caption{Visual and qualitative comparisons across multiple scenes. We present the rendering results of our method using 128 reflected rays (\textbf{Ours-128}) and 32 reflected rays (\textbf{Ours-32}), in comparison with vanilla Neural Radiosity (\textbf{NR}), an equal-time Monte Carlo path tracer with 4 spp and Oidn denoising (\textbf{Oidn}), and a 16 spp path tracer (\textbf{PT}). \textbf{MAPE} is reported with respect to the reference (path traced with 100{,}000 spp). The metric shown below each row indicates the overall MAPE, while the value in the bottom-right corner of each region shows its local error.}
    \label{fig:compare}
\end{figure*}

We compare our model with vanilla Neural Radiosity (NR)~\cite{hadadan2021neural}, equal-time path tracer (PT), and PT with Monte Carlo denoiser (Oidn)~\cite{OpenImageDenoise}. For our method, we show the results of tracing $T=128$ and $T=32$ secondary rays. For NR, the MLP of the model consists of 4 hidden layers, each has 256 neurons, while the configuration of feature grid is the same as the diffuse feature grid $\mathbf{v}_{dif}$ of ours. For all the comparisons, we evaluate image quality with mean absolute percentage error (MAPE) as the error metric. As shown in Figure~\ref{fig:compare}, our method with different settings outperforms NR and PT, and is comparable with Oidn. Qualitative results show that reducing the amount of reflected rays during rendering only slightly compromises the image quality. For NR, the reconstruction quality greatly deteriorates when there is high-frequency reflection contents in glossy region. Though the image quality of Oidn is comparable to ours, the denoised path tracing results show apparent flickering artifacts during interactive rendering, while our method shows much better temporal stability, and capture high-frequency details better. Please refer to our supplementary video for interactive rendering results.

The runtime performance is shown in Table~\ref{tbl:time}. We calculated the average time overhead per frame. For each method and each scene, we rendered 10 frames as warm-up stage, then calculated the average rendering time of 100 frames. Results show that reducing the number of reflected rays effectively reduced the time cost. Although extra modules are introduced into our pipeline, the rendering time cost of our method is still comparable to NR due to smaller MLP size.

% Performance

\begin{table}
\caption{Performance evaluation on our method with 128 reflected rays (\textbf{Ours-128}) and 32 reflected rays (\textbf{Ours-32}), in comparison with vanilla Neural Radiosity (\textbf{NR}), a 4-spp Monte Carlo path tracer with Oidn denoising (\textbf{Oidn}), and a 16-spp path tracer (\textbf{PT}). Per-frame time cost (in milliseconds) and image quality (MAPE) are included.}
\begin{tabular}{lrrrrr}
\toprule
\mc{Time (ms)\\ MAPE} & Ours-128 & Ours-32 & NR & Oidn & PT \\
\midrule
bathroom & \mc{82\\0.096} & \mc{63\\0.096} & \mc{65\\0.134} & \mc{80\\0.128} & \mc{198\\0.518} \\
\hline
\noalign{\vskip 1pt}
cornell-box & \mc{49\\0.057} & \mc{37\\0.059} & \mc{33\\0.073} & \mc{41\\0.070} & \mc{97\\0.325} \\
\hline
\noalign{\vskip 1pt}
kitchen & \mc{131\\0.068} & \mc{116\\0.069} & \mc{136\\0.124} & \mc{109\\0.084} & \mc{182\\0.440} \\
\hline
\noalign{\vskip 1pt}
living-room & \mc{72\\0.090} & \mc{53\\0.092} & \mc{45\\0.109} & \mc{53\\0.116} & \mc{123\\0.577} \\
\hline
\noalign{\vskip 1pt}
veach-ajar & \mc{49\\0.101} & \mc{40\\0.105} & \mc{38\\0.141} & \mc{41\\0.135} & \mc{90\\0.178} \\
\bottomrule
\end{tabular}
\label{tbl:time}
\end{table}

The training time of our method ranges from 1 to 5 hours per scene, depending on scene complexity. This is approximately 60\% longer than vanilla Neural Radiosity, as we only optimize the CUDA kernel for the forward pass used during rendering, while the training pipeline remains implemented in pure PyTorch. Our model requires 104MB of parameters per scene, compared to 44MB for Neural Radiosity. Although this results in additional storage overhead for checkpoints, the increase can be negligible for modern GPU.

We did not compare our method with other state-of-the-art neural global illumination approaches~\cite{zheng2023nelt, ren2024lightformer, zheng2024neural}, as few of them explicitly target highly glossy effects, and our cone-encoding strategy is orthogonal to their design. Moreover, although some methods have demonstrated glossy illumination results~\cite{muller2021real}, they do not query the network at the primary intersection, which often leads to noisy renderings. As a result, a direct comparison under our evaluation setting would be less meaningful, which focuses on primary-surface predictions with high-frequency view-dependent effects.

\subsection{Validation}

\begin{figure}
    \centering
    \includegraphics[width=\linewidth]{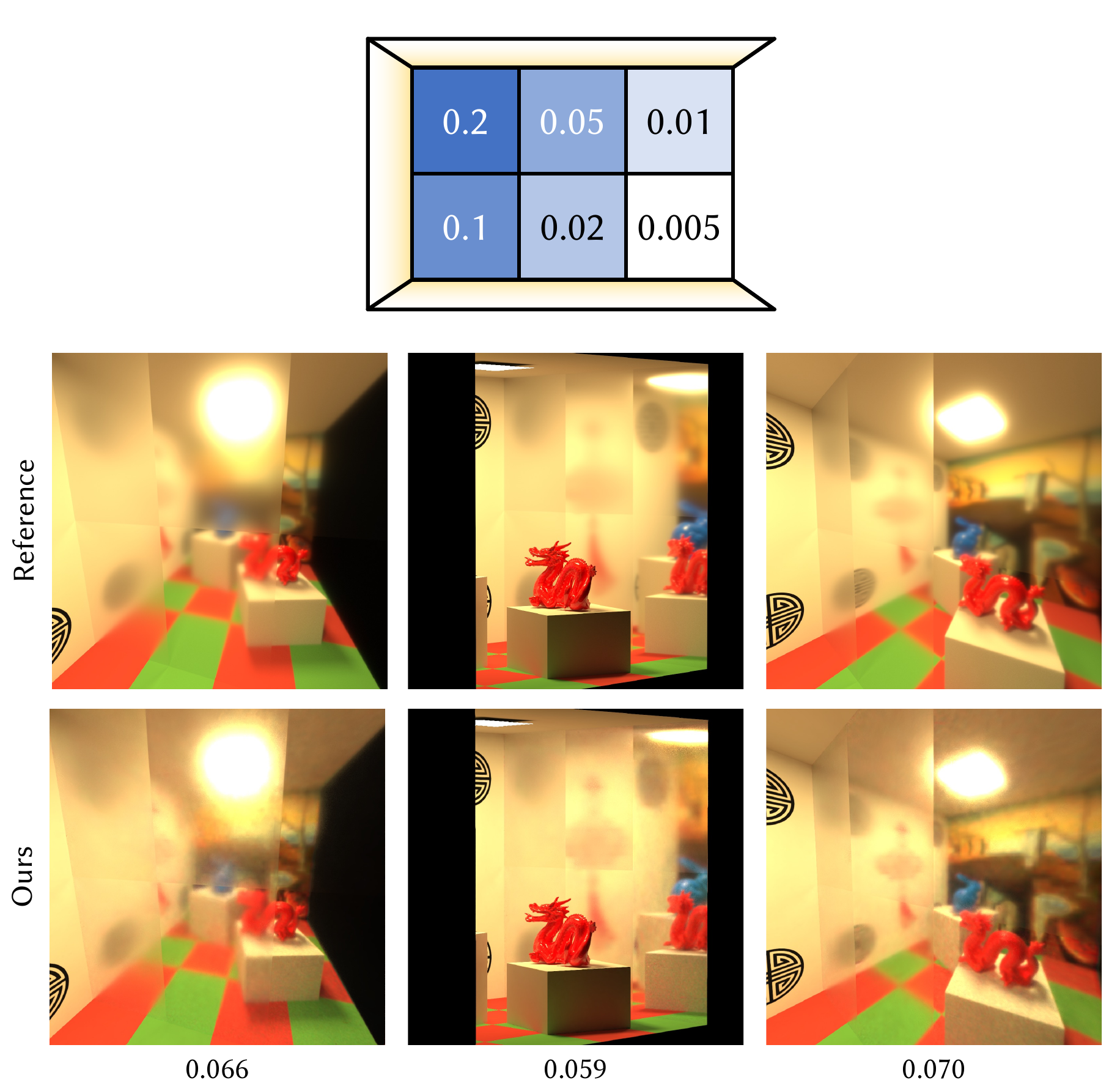}
    \caption{Spatially varying roughness on the wall of the Cornell box scene. The first row shows the references rendered from different camera poses, while the second row presents the results produced by our method, with MAPE values included.}
    \label{fig:roughness}
\end{figure}

% Varying roughness
To demonstrate the generalization capability of our model across different surface roughness levels, we designed a scene where the roughness varies spatially on a single object—the right wall of the Cornell Box. The roughness ranges from 0.2 to 0.005 across different rectangular regions (see Figure~\ref{fig:roughness}). Results show that our glossy model successfully reconstructs corresponding reflected contents across varying degrees of roughness, indicating its robustness to surface variation.

\begin{figure}
    \centering
    \includegraphics[width=\linewidth]{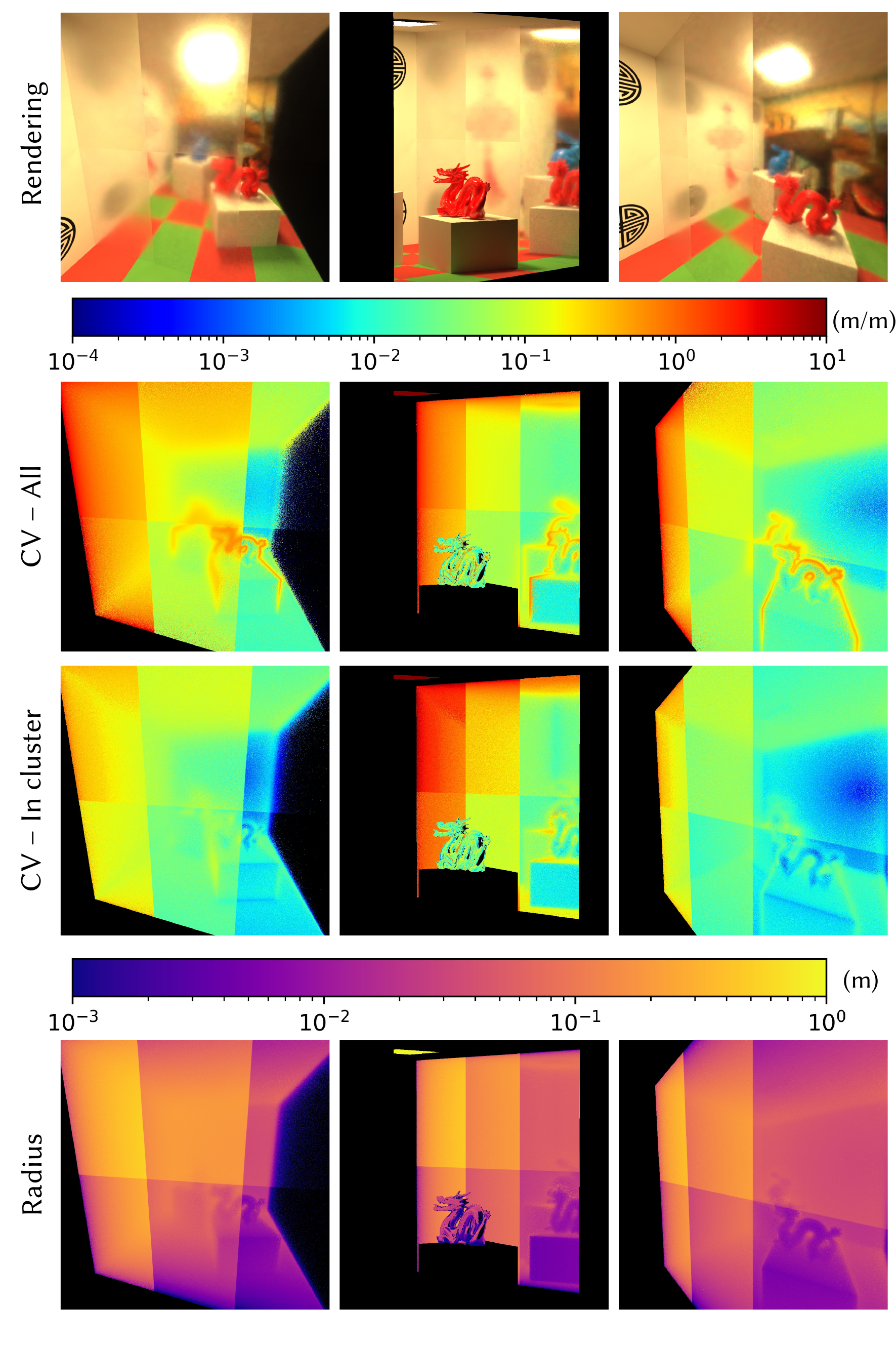}
    \caption{Visualization of the coefficient of variation (CV) in reflected ray marching distances. The first row shows rendered images from different camera views using our method. The second row visualizes the CV computed from all $T = 128$ reflected rays. The third row shows the weighted average CV within each cluster under the clustering approximation. The last row displays the average query radius used by the network. Clustering approximation significantly reduces intra-cluster variance, enabling more stable and efficient network queries.}
    \label{fig:vis_t}
\end{figure}

To demonstrate the necessity of clustering approximation, we visualize the coefficient of variation (CV), defined as the standard deviation divided by the mean, of the marching distances of reflected rays (see Figure~\ref{fig:vis_t}). The first row shows results from our method under different camera views. The second row presents the overall CV computed from all $T = 128$ reflected rays per surface point, while the third row shows the weighted average CV within each cluster after applying the clustering approximation. The results indicate that reflected ray depths exhibit high variance, particularly around the silhouettes of reflected objects. Clustering significantly reduces this variance within each group, enabling more structured and localized network queries. The last row displays the average query radius per point, which varies smoothly and aligns with the projected area of the cone–surface intersection.

\begin{figure}
    \centering
    \includegraphics[width=\linewidth]{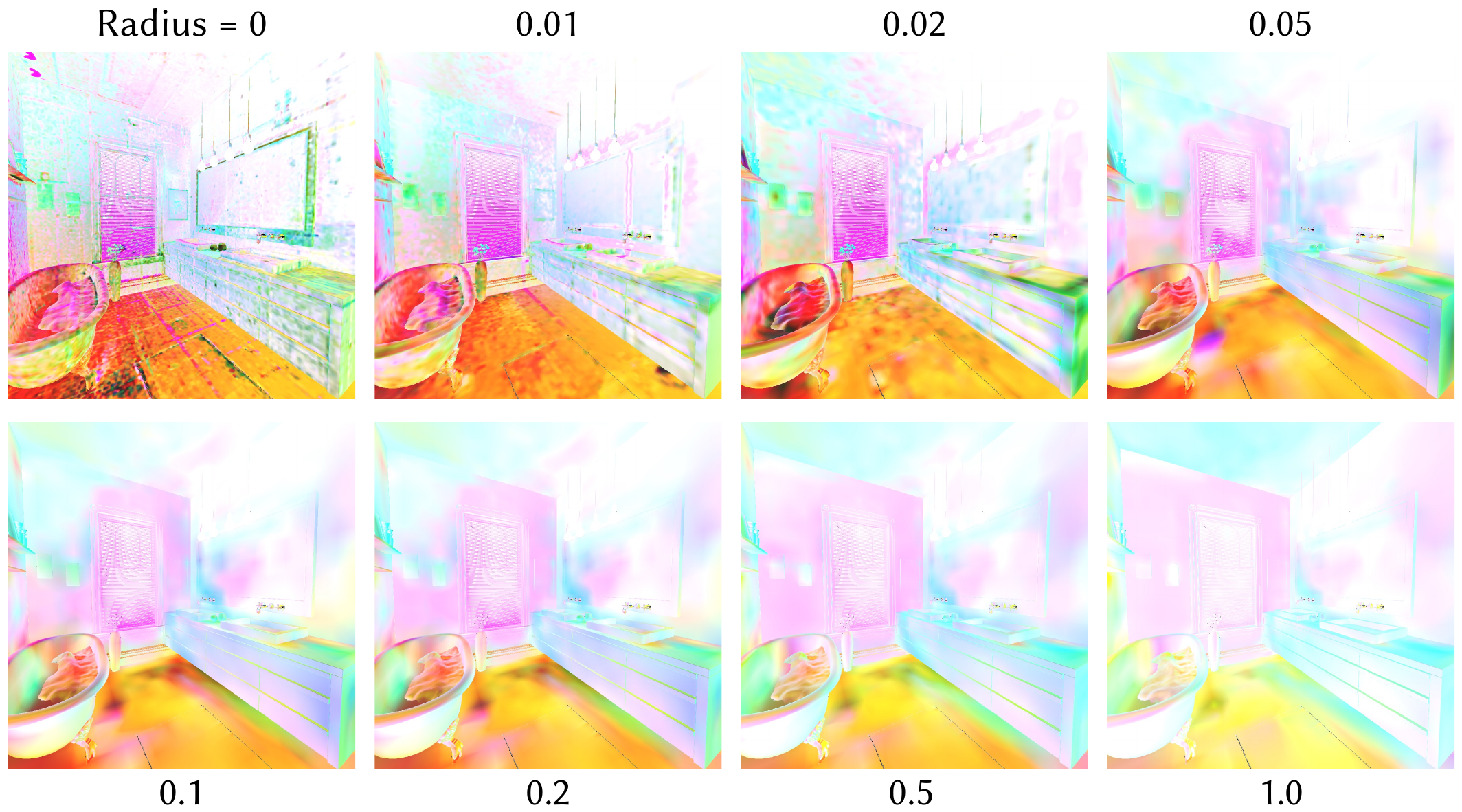}
    \caption{Visualization of the output from the glossy model. Each image displays the 3-dimensional output of the glossy model. The numbers shown above and below the images indicate the radius used when querying the glossy model. Note that these outputs serve merely as inputs to the modulation network, and therefore, their colors appear significantly different from the final rendering results.}
    \label{fig:visualize}
\end{figure}

% Visualization of glossy model
To illustrate the behavior of the glossy model, we directly evaluate the reflection model at the primary intersection point using varying query radii (see Figure~\ref{fig:visualize}). The visualization shows that when the radius is set to zero, the glossy model captures high-frequency details using the finest level of the feature grid. As the radius increases, features from multiple resolution levels are aggregated, resulting in a progressively more convolved feature distribution with a larger effective kernel size. This illustrates how the model smoothly transitions from fine to coarse features depending on the spatial support.

\begin{figure}
    \centering
    \includegraphics[width=\linewidth]{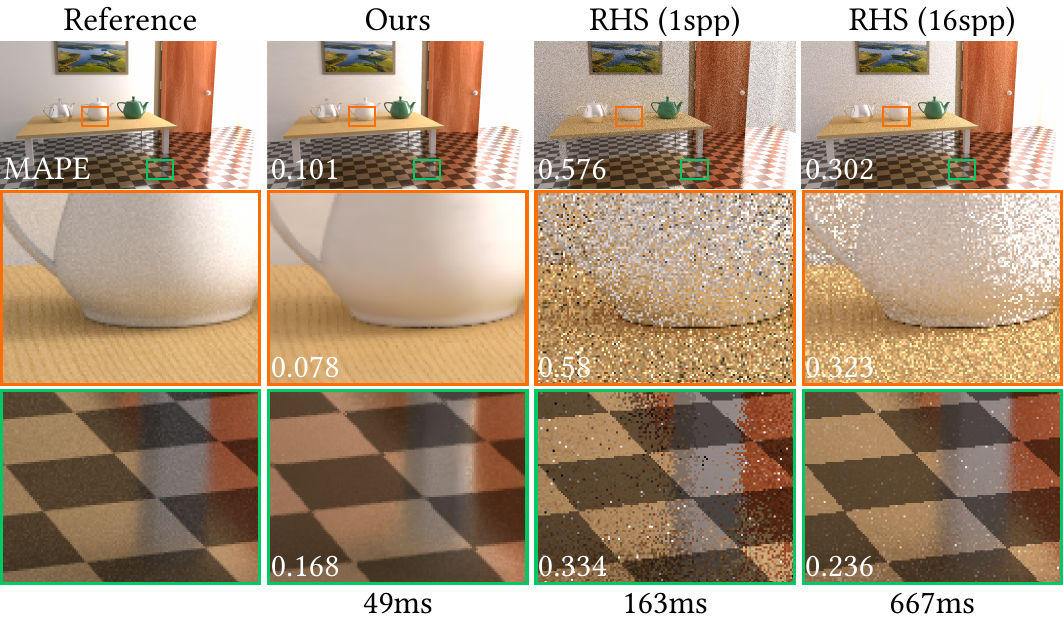}
    \caption{Comparison between our method and the RHS variant of Neural Radiosity. We render images using our method (\textbf{Ours}), the right-hand-side variant of Neural Radiosity with 1 sample per pixel (\textbf{RHS (1spp)}), and with 16 samples per pixel (\textbf{RHS (16spp)}). \textbf{MAPE} is reported in the bottom-left corner of both the full image and the cropped region. Per-frame time cost is reported below the images.}
    \label{fig:rhs}
\end{figure}

% RHS vs NCR
NR was originally proposed as an offline rendering method, with limited emphasis on performance. It supports rendering by querying the network at either LHS (primary intersection) or RHS (secondary intersection) of the radiosity equation. In this experiment, we compare our method against the RHS variant of NR, which is also capable of rendering glossy reflections (see Figure~\ref{fig:rhs}). Results show that our method produces images with less noise and achieves shorter rendering times compared to the RHS variant.

\begin{figure}
    \centering
    \includegraphics[width=\linewidth]{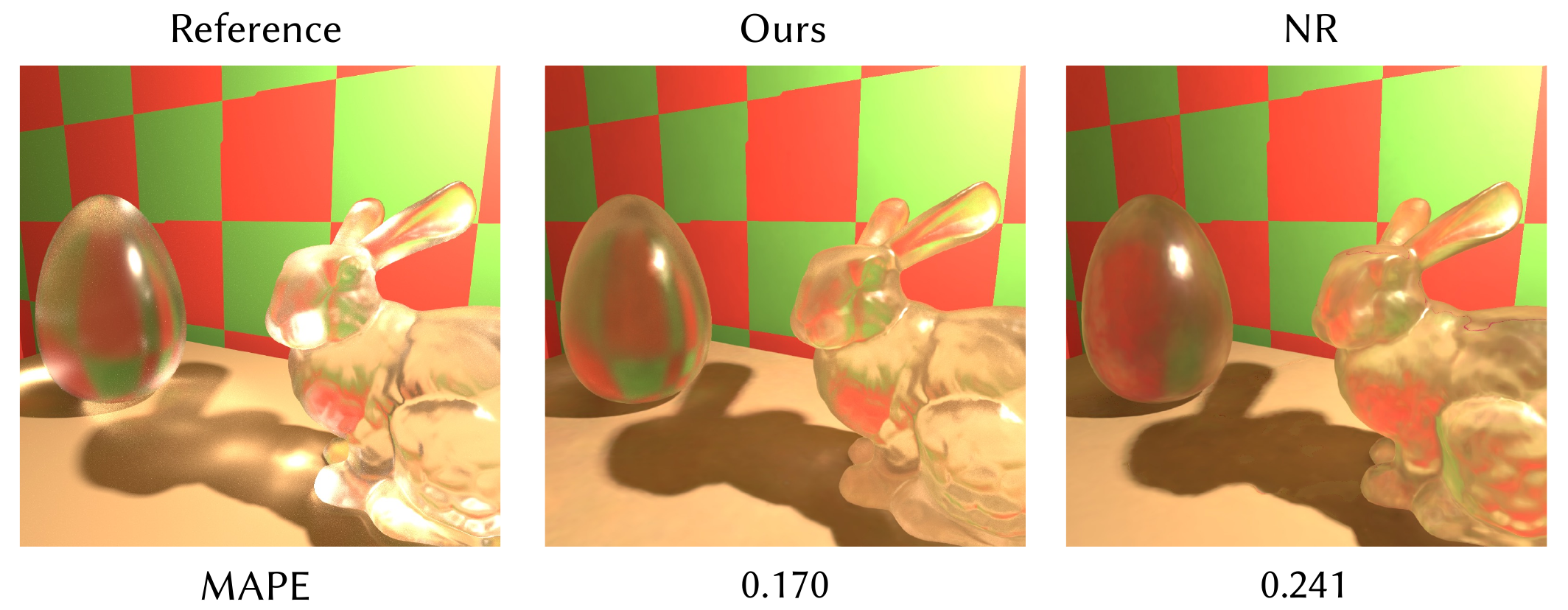}
    \caption{Comparison of glossy refraction from rough dielectric materials. We present the reference (left), ours (middle), and NR (right), with MAPE included. Our method more accurately reconstructs the refracted and reflected compared to NR, though neither can capture the caustic effects.}
    \label{fig:dielectric}
\end{figure}

% Rough dielectric materials
Beyond glossy reflections, our method also demonstrates the potential to model glossy refractions (see Figure~\ref{fig:dielectric}). For glossy dielectric materials, we trace both reflection and refraction cones separately. The rendered results show that our method outperforms NR in reconstructing the geometry of refracted and reflected content.

However, neither our method nor NR is able to reproduce caustic effects on glossy dielectric materials, which significantly degrade image quality. This limitation primarily arises from the inability to sample light sources effectively after undergoing complex light transport within dielectric materials. One possible solution is to incorporate bidirectional sampling strategies~\cite{su2025proxy} during training. Since sampling strategies are orthogonal to our core method, we leave this direction for future work.

\begin{figure}
    \centering
    \includegraphics[width=\linewidth]{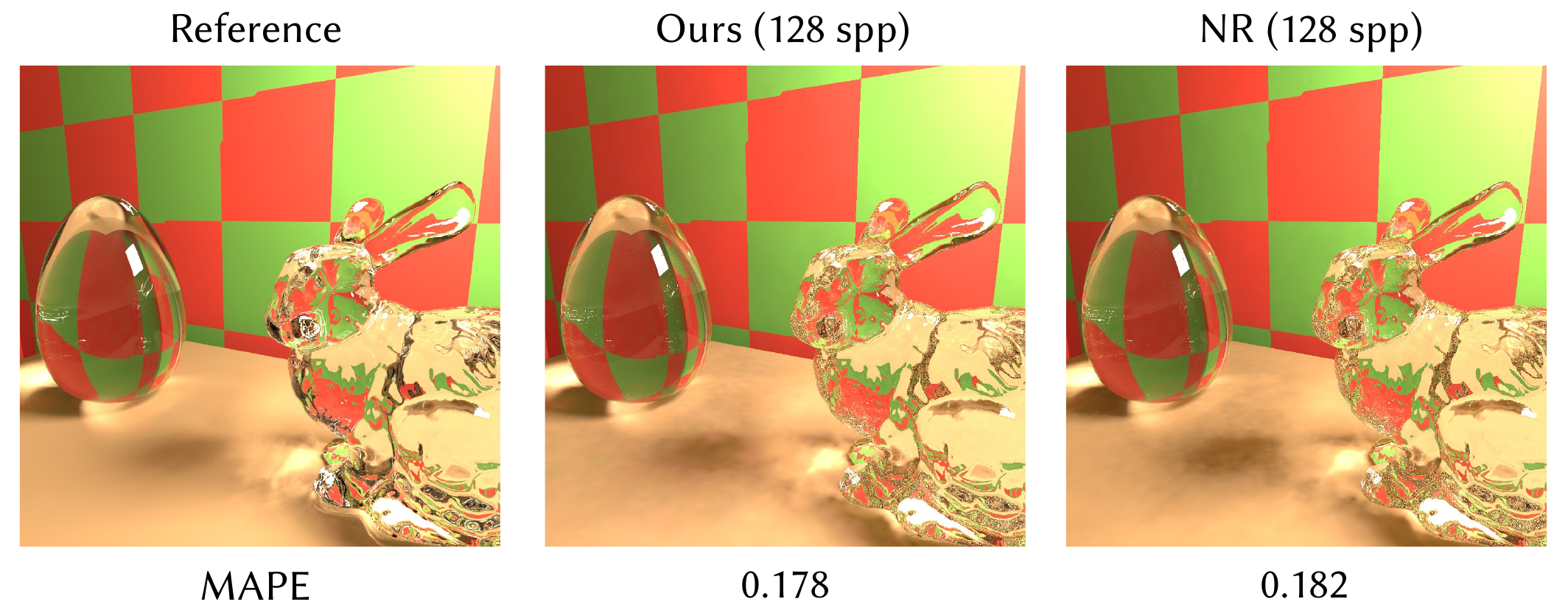}
    \caption{Caustic effects on specular dielectric materials. We present the reference (left), ours (middle), and Neural Radiosity (NR, right). MAPE with respect to the reference is included. Both images of ours and NR are rendered with 128 spp, to reduce the sampling noise of reflection or refraction at specular dielectric surfaces. Our method achieves similar quality in reconstructing caustic effects under specular refractions.}
    \label{fig:spec-diel}
\end{figure}

In contrast, the caustic effects produced by pure dielectric materials can be successfully captured by both our method and NR (See Figure~\ref{fig:spec-diel}). This is because the radiance from RHS supervision is not evaluated until the first non-specular surface is encountered. In the case of caustics, this surface typically lies on the light source corresponding to the bright spot region.

\subsection{Ablation Study}

\begin{figure}
    \centering
    \includegraphics[width=\linewidth]{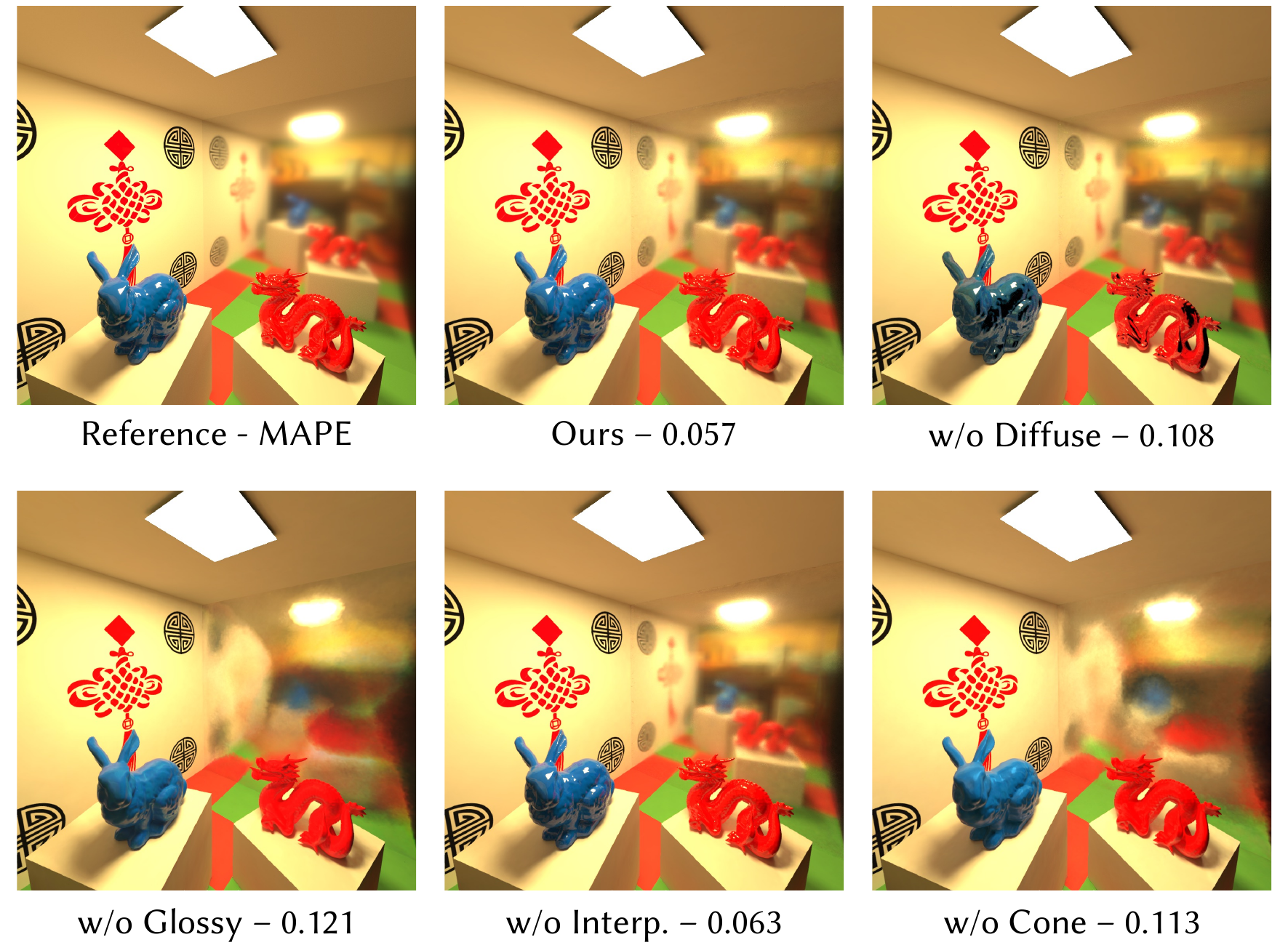}
    \caption{Ablation study on key model components in the \emph{cornell-box} scene. We compare the reference image, our full model (\textbf{Ours}), and four ablated variants: without the diffuse network (\textbf{w/o Diffuse}), without the glossy network (\textbf{w/o Glossy}), without layer interpolation (\textbf{w/o Interp.}), and without cone encoding (\textbf{w/o Cone}). \textbf{MAPE} is reported below.}
    \label{fig:ablation}
\end{figure}

We conducted an ablation study on the \emph{cornell-box} scene to evaluate the effectiveness of key components in our model (see Figure~\ref{fig:ablation}). Specifically, we trained variants of our model with the following components removed: the diffuse network (\textbf{w/o Diffuse}), the glossy network (\textbf{w/o Glossy}), layer interpolation (\textbf{w/o Interp.}), and cone encoding (\textbf{w/o Cone}).

In the \textbf{w/o Diffuse} setting, the model relies solely on the glossy network, with weighted outputs used directly as predictions, bypassing the modulation network. In the \textbf{w/o Glossy} variant, the diffuse prediction is used for all non-specular pixels. \textbf{w/o Interp.} disables layer interpolation by applying a mean reduction over all resolution levels. In the \textbf{w/o Cone} case, ray–surface intersections along the specular direction are directly fed into the glossy network, without using cone encoding.

Experimental results show that the removal of any single component leads to a noticeable degradation in image quality, highlighting the necessity of each module.

\section{Conclusion and Future Work}

%We present Neural Cone Radiosity, a novel approach for modeling global illumination effects on glossy surfaces while maintaining interactive rendering performance. To achieve this, we introduce a cone-based encoding for glossy reflections and employ feature grid representations. A clustering-based approximation is used to decompose complex cone–surface intersections into spherical regions that can be efficiently processed via single network queries. Furthermore, we design a dual-branch architecture to separately modulate diffuse and glossy components, enabling more accurate modeling of both rough and smooth reflections.

In this paper, we have presented a novel approach for modeling global illumination effects on glossy surfaces, enabling accurate and efficient handling of glossy material. Our approach addresses the spatial continuity and integration challenges inherent in glossy transport, demonstrating that neural representations can move beyond point-based approximations toward scale-aware, physically consistent formulations.
Experimental validations have demonstrated that our method yields superior image quality on both glossy and non-glossy surfaces.

More broadly, our framework illustrates how pre-filtered neural operators and clustering-based integration strategies can systematically reduce the complexity of global illumination while preserving high visual fidelity. These contributions highlight a pathway for unifying efficiency, scalability, and physical realism in neural rendering, which may bring us closer to a general-purpose solution for real-time global illumination in complex scenes.

Despite its effectiveness, our method still faces certain limitations in terms of generalization. Specifically, the model must be trained from scratch for each scene configuration. While dynamic neural radiosity~\cite{su2024dynamic} offers a potential solution for handling variations across multiple dimensions, it is not yet incorporated into our framework. Additionally, our approach struggles to accurately model refraction effects in transparent materials due to the complexity introduced by multi-bounce, highly directional light transport.

As part of future work, we plan to integrate our cone encoding technique with generalizable neural rendering frameworks~\cite{zeng2025renderformer, zheng2023nelt}. We believe this represents a promising step toward making neural rendering methods viable for industrial-scale applications.

\bibliographystyle{abbrv-doi}
\bibliography{references}

\onecolumn

%\appendix

\end{document}